\documentclass[sigconf]{acmart}

\AtBeginDocument{%
  }

\setcopyright{cc}
\setcctype{by-nc-nd}
\copyrightyear{2027}
\acmYear{2027}
\acmDOI{10.1145/3840404.3848292}

\acmConference[ASP-DAC '27]{the 32nd Asia and South Pacific Design Automation Conference}{January 25--28, 2027}{Tokyo, Japan}
\acmISBN{979-8-4007-2923-2}

\usepackage{algorithm}
\usepackage{algorithmic}

\usepackage{array}
\usepackage{multirow}
\newcolumntype{C}[1]{>{\centering\arraybackslash}p{#1}}
\usepackage{booktabs}
\usepackage{makecell}
\usepackage{tabularx}

\begin{document}

\title{PaFeDSE: Pareto-Feedback-Guided Microarchitecture Design Space Exploration for Efficient Pareto Search}

\author{Joonyong Park}
\affiliation{
    \department{School of Electrical Engineering}
    \institution{KAIST}
    \city{Daejeon}
    \country{Republic of Korea}}
\email{wndy2001@kaist.ac.kr}

\author{Hyojeong Lee}
\affiliation{
    \department{School of Electrical Engineering}
    \institution{KAIST}
    \city{Daejeon}
    \country{Republic of Korea}}
\email{hjlee8877@kaist.ac.kr}

\author{Sejeong Yang}
\affiliation{
    \department{School of Electrical Engineering}
    \institution{KAIST}
    \city{Daejeon}
    \country{Republic of Korea}}
\email{02yangsj@kaist.ac.kr}

\author{Joo-Young Kim}
\affiliation{
    \department{School of Electrical Engineering}
    \institution{KAIST}
    \city{Daejeon}
    \country{Republic of Korea}}
\email{jooyoung1203@kaist.ac.kr}

%%%%%%%%%%%%%%%%%%%%%%%%%%%%%%%%%%%%%%%%%%%%%%%%%%%%%%%%%%%%%%%%%%%%%%%%%%%
%%%%%%%%%%%%%%%%%%%%%%%%%%%%%%%%%%%%%%%%%%%%%%%%%%%%%%%%%%%%%%%%%%%%%%%%%%%

\begin{abstract}
Efficient microarchitecture design space exploration (DSE) is critical for optimizing performance, power, and area trade-offs in modern out-of-order processors.
While machine-learning-based DSE significantly reduces the cost of exhaustively evaluating large design spaces, existing methods still waste expensive simulations on Pareto-miss candidates that do not improve the Pareto frontier and repeatedly process redundant candidates during iterative exploration.
In this paper, we propose \textit{PaFeDSE}, a Pareto-feedback-guided microarchitecture DSE framework for efficient Pareto search.
PaFeDSE leverages Pareto-hit feedback to prioritize candidates likely to improve the frontier and Pareto-miss feedback to suppress exploration of unpromising local regions, while filtering redundant candidates that remain low priority across iterations.
Evaluated on SPEC CPU 2017 workloads, PaFeDSE requires up to 49.0\% fewer simulations than the strongest prior baseline to reach target hypervolume levels, while achieving a 7.1$\times$ reduction in exploration overhead through adaptive candidate filtering.
\end{abstract}

%%%%%%%%%%%%%%%%%%%%%%%%%%%%%%%%%%%%%%%%%%%%%%%%%%%%%%%%%%%%%%%%%%%%%%%%%%%
%%%%%%%%%%%%%%%%%%%%%%%%%%%%%%%%%%%%%%%%%%%%%%%%%%%%%%%%%%%%%%%%%%%%%%%%%%%




\maketitle

%%%%%%%%%%%%%%%%%%%%%%%%%%%%%%%%%%%%%%%%%%%%%%%%%%%%%%%%%%%%%%%%%%%%%%%%%%%
%%%%%%%%%%%%%%%%%%%%%%%%%%%%%%%%%%%%%%%%%%%%%%%%%%%%%%%%%%%%%%%%%%%%%%%%%%%

\section{Introduction}

The growing complexity of modern out-of-order processors has made microarchitecture design space exploration (DSE) essential for identifying high-quality performance, power, and area (PPA) trade-offs.
The goal of microarchitecture DSE is to identify Pareto-optimal configurations that balance these competing objectives within a design space, where each design point represents a specific configuration of microarchitectural parameters.
However, exhaustively evaluating all design points is practically infeasible because each evaluation requires costly architectural simulation and PPA estimation.
Therefore, recent DSE methods have increasingly adopted machine learning (ML)-based surrogate models to approximate costly design evaluations~\cite{modse_dac,modse_tcad,boomexplorer_tdaes,boomexplorer_iccad,grl,metadse,actboost,swiftorexact}.
By learning the relationship between microarchitectural parameters and PPA metrics from a small set of evaluated designs, a surrogate model can rapidly estimate the PPA of unevaluated candidates without performing costly evaluations for every design point.

\begin{figure}[t]
    \centering
    \includegraphics[
        width=0.99\linewidth,
        trim=23 6 33 11,
        clip
    ]{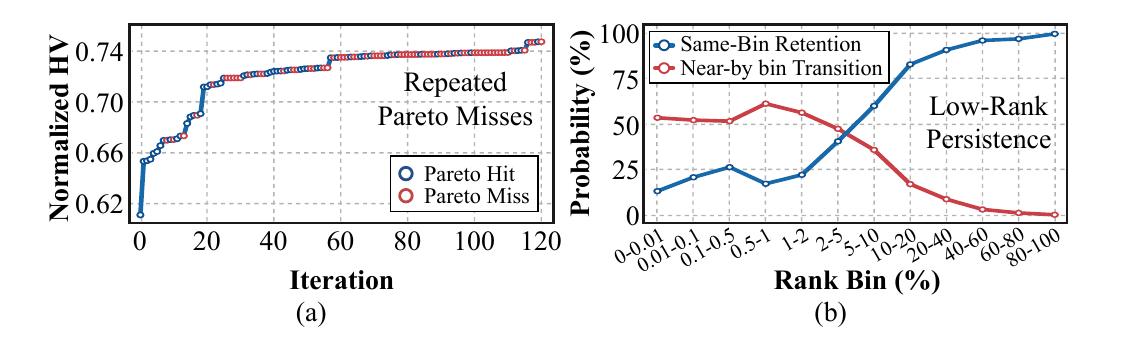}
    \caption{Analysis of MoDSE on 638.imagick\_s. (a) Normalized HV curve across iterations. (b) Retention and transition probabilities across rank bins between adjacent iterations.}
    \Description{Two plots showing HV improvement with frequent Pareto misses and rank-bin transition probabilities indicating that low-priority candidates tend to remain low-priority across iterations.}
    \label{figure1}
\end{figure}

Most ML-based DSE methods~\cite{modse_dac,modse_tcad,boomexplorer_tdaes,boomexplorer_iccad,actboost,swiftorexact} follow an iterative process in which each iteration consists of three main steps:
1) predicting the PPA of all unevaluated candidate design points using a surrogate model, 
2) scoring the candidates with an acquisition function and selecting the candidate most likely to improve the current observed Pareto frontier, and
3) evaluating the selected candidate with a cycle-accurate simulator and updating the observed Pareto frontier.
The evaluated design point is then added to the training set for the next iteration.
Representative DSE methods adopt various surrogate models, including Gaussian process (GP) models, ensemble regressors, and neural predictors~\cite{modse_dac,boomexplorer_iccad,actboost}.
Many of them use hypervolume (HV)-based acquisition functions, such as hypervolume improvement (HVI) and expected hypervolume improvement (EHVI)~\cite{hvi,ehvi}.
This iterative DSE loop enables exploration within a limited simulation budget by evaluating only selected candidates rather than the entire design space.
Nevertheless, existing ML-based DSE methods still fall short in Pareto-oriented search efficiency, leaving two key limitations.

\textit{First}, \textbf{existing ML-based DSE methods waste a large portion of iterations on Pareto-miss selections.} 
We call a selected candidate a \emph{Pareto hit} if its evaluated PPA improves the hypervolume of the current observed Pareto set ($\Delta \mathrm{HV}>\varepsilon$), and a \emph{Pareto miss} otherwise.
Because each DSE iteration still requires expensive simulation of the selected design point, DSE efficiency largely depends on whether the selected candidate meaningfully improves the current observed Pareto frontier.
However, as shown in Figure~\ref{figure1}(a), MoDSE~\cite{modse_dac}, a state-of-the-art ML-based DSE method, spends many iterations on Pareto-miss selections ($\varepsilon=0.0001$).
Across the evaluated SPEC CPU 2017 workloads~\cite{speccpu2017}, the maximum Pareto-miss rate of MoDSE reaches 85.47\%.
This inefficiency stems from surrogate models limited to PPA prediction and the lack of explicit Pareto-feedback utilization.
Without such feedback, previously evaluated design points are used only as PPA training samples, while their Pareto-hit or Pareto-miss outcomes are not leveraged to promote candidates likely to become Pareto hits or suppress regions that repeatedly lead to Pareto misses.

\textit{Second}, \textbf{iterative DSE redundantly processes persistently low-priority candidates.}
Modern CPU microarchitectures involve a growing number of tunable parameters, leading to massive design spaces.
Since each DSE iteration typically performs surrogate prediction and acquisition scoring over all unevaluated candidates, this process introduces non-negligible exploration overhead as the design space expands.
However, repeatedly processing all unevaluated candidates is largely unnecessary.
Figure~\ref{figure1}(b) groups candidates by acquisition-score rank percentile and tracks whether they remain in the same rank bin in the next iteration.
The high retention probability in low-rank bins shows that candidates in these bins are likely to remain low-priority in subsequent iterations.
Nevertheless, existing methods continue to reprocess these low-priority candidates in every iteration, incurring redundant candidate-processing overhead for designs that are unlikely to affect the acquisition outcome.

To address these limitations, we propose \textit{PaFeDSE}, a Pareto-feedback-guided microarchitecture DSE framework for efficient Pareto search.
PaFeDSE improves DSE efficiency by exploiting Pareto feedback observed during iterative exploration, where Pareto hits guide reliable candidate selection and Pareto misses guide locality-aware penalization. 
Moreover, PaFeDSE leverages inter-iteration consistency to reduce redundant candidate processing. The main contributions of this work are summarized as follows:

\begin{itemize}
\item We introduce \textit{Pareto-Hit-Aware Transformer Surrogate} (\textit{HitFormer}), a Transformer-based surrogate model for joint prediction of PPA and Pareto-context-aware hit score to improve the reliability of HVI-based candidate selection.

\item We propose \textit{Locality-Aware Pareto-Miss Penalty Graph} (\textit{MissGraph}), a graph-based memory mechanism that records previous Pareto misses and penalizes nearby unpromising candidates using PPA-sensitive design distances.

\item We present \textit{Temporal-Consistency-Guided Filtering} (\textit{TCFilter}), which exploits inter-iteration acquisition-score consistency to reduce redundant candidate processing while preserving observed Pareto-set quality.

\item We evaluate \textit{PaFeDSE} on SPEC CPU 2017, demonstrating that it reduces the number of simulations required to reach high hypervolume levels by up to 49.0\% compared with the prior state-of-the-art ML-based DSE method, while adaptive filtering achieves a 7.1$\times$ reduction in exploration overhead.
\end{itemize}

%%%%%%%%%%%%%%%%%%%%%%%%%%%%%%%%%%%%%%%%%%%%%%%%%%%%%%%%%%%%%%%%%%%%%%%%%%%
%%%%%%%%%%%%%%%%%%%%%%%%%%%%%%%%%%%%%%%%%%%%%%%%%%%%%%%%%%%%%%%%%%%%%%%%%%%
\section{Background}
\subsection{DSE Problem Formulation}
Let $D$ denote the finite design space, where each point $x \in D$ is a microarchitectural parameter configuration.
Each point is evaluated by architectural simulation and PPA estimation, which maps $x$ to an objective vector $f(x)$.
Accordingly, we formulate microarchitecture DSE as a multi-objective minimization problem:
\begin{equation}
\min_{x \in D} f(x) =
\{ -\mathrm{Performance}(x), \mathrm{Power}(x), \mathrm{Area}(x) \}.
\end{equation}
Because exhaustively evaluating all points in $D$ is prohibitively expensive, DSE methods typically adopt surrogate-assisted iterative exploration.
At each iteration, a surrogate model approximates $f(x)$ for unevaluated candidates, and an acquisition function selects the candidate most likely to improve the observed Pareto frontier for evaluation.
This process aims to identify high-quality PPA trade-offs while evaluating only a small fraction of the design space.

\subsection{Pareto Optimality and Hypervolume}
\label{sec:background_hv}
In an $n$-objective minimization problem, a design point $x'$ dominates another design point $x$, denoted as $x' \prec x$, if $x'$ is no worse than $x$ in all objectives and strictly better in at least one objective:
\begin{equation}
\left( \forall i \in [n],\; f_i(x') \leq f_i(x) \right)
\land
\left( \exists j \in [n],\; f_j(x') < f_j(x) \right),
\end{equation}
where $[n]=\{1,\ldots,n\}$. 
Based on this dominance relation, the Pareto set $\Omega$ is defined as the set of design points in $D$ that are not dominated by any other design point in $D$:
\begin{equation}
\Omega =
\left\{
x^* \in D \mid
\forall x \in D,\; x \not\prec x^*
\right\}.
\end{equation}

HV measures the objective-space volume dominated by the Pareto set $\Omega$ with respect to a reference point $\mathbf{r}$~\cite{hv}.
In a minimization problem, the reference point $\mathbf{r}$ is chosen to be worse than the Pareto points in all objectives and serves as the upper bound of the dominated objective-space region.
Let $\lambda_n(\cdot)$ denote the $n$-dimensional Lebesgue measure.
The HV of $\Omega$ is defined as
\begin{equation}
\mathrm{HV}(\Omega) =
\lambda_n
\left(
\bigcup_{x \in \Omega}
\prod_{i=1}^{n}
[f_i(x), r_i]
\right).
\end{equation}
In iterative DSE, let $\Omega_{\mathrm{obs}}$ denote the current observed Pareto set among the evaluated design points.
Given a candidate design point $x$, hypervolume improvement (HVI)~\cite{hvi} measures the increase in HV obtained by adding $x$ to $\Omega_{\mathrm{obs}}$:
\begin{equation}
\mathrm{HVI}(x \mid \Omega_{\mathrm{obs}}) =
\mathrm{HV}(\Omega_{\mathrm{obs}} \cup \{x\}) - \mathrm{HV}(\Omega_{\mathrm{obs}}).
\end{equation}

%%%%%%%%%%%%%%%%%%%%%%%%%%%%%%%%%%%%%%%%%%%%%%%%%%%%%%%%%%%%%%%%%%%%%%%%%%%
%%%%%%%%%%%%%%%%%%%%%%%%%%%%%%%%%%%%%%%%%%%%%%%%%%%%%%%%%%%%%%%%%%%%%%%%%%%
\section{PaFeDSE Framework}
\subsection{Overview of PaFeDSE}

\begin{figure}[t]
    \centering
    \includegraphics[
        width=\linewidth,
        trim=20 9 20 8,
        clip
    ]{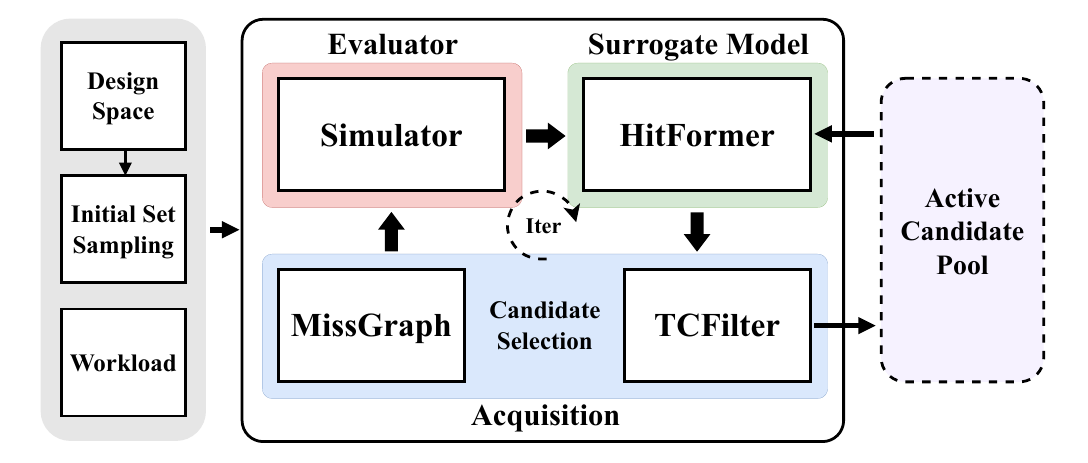}
    \caption{Overview of PaFeDSE.}
    \Description{Block diagram of the PaFeDSE framework, showing the iterative design evaluation and exploration process involving HitFormer, MissGraph, TCFilter, and the active candidate pool.}
    \label{figure2}
\end{figure}

\begin{algorithm}[t]
\caption{PaFeDSE exploration framework}
\label{alg:pafedse}
\footnotesize
\begin{algorithmic}[1]
\REQUIRE Design space $\mathcal{D}$, workload $w$, maximum iteration number $Iter_{\max}$, initial set size $n_0$
\ENSURE Observed Pareto set $\Omega$

\STATE $X \leftarrow \mathrm{SensitivityAwareInitialSetSampling}(\mathcal{D}, n_0)$
\STATE $S \leftarrow \{(x, \mathrm{simulate}(x,w)) \mid x \in X\}$, $U \leftarrow \mathcal{D} \setminus X$
\STATE $\Omega \leftarrow \mathrm{GetCurrentParetoSet}(S)$
\STATE $G \leftarrow \mathrm{BuildMissGraph}(\mathcal{D}, S)$
\STATE $A \leftarrow U$
\FOR{$iter \leftarrow 1$ to $Iter_{\max}$}
    \STATE $M \leftarrow \mathrm{TrainHitFormer}(S)$
    \FORALL{$x \in A$}
        \STATE $\hat{y}(x), p_{\mathrm{hit}}(x) \leftarrow M(x,\Omega)$
        \STATE $g(x) \leftarrow \mathrm{GetMissGraphFactor}(G,x)$
        \STATE $score[x] \leftarrow \mathrm{HVI}(\hat{y}(x)\mid\Omega)\cdot p_{\mathrm{hit}}(x)\cdot g(x)$
    \ENDFOR
    % \FORALL{$x \in A$}
    %     \STATE $\hat{y}, p_{\mathrm{hit}} \leftarrow M(x,\Omega)$
    %     \STATE $g \leftarrow \mathrm{GetMissGraphFactor}(G,x)$
    %     \STATE $score[x] \leftarrow \mathrm{HVI}(\hat{y} \mid \Omega) \cdot p_{\mathrm{hit}} \cdot g$
    % \ENDFOR
    \STATE $x^{*} \leftarrow \arg\max_{x \in A} score[x]$
    \STATE $y^{*} \leftarrow \mathrm{Simulate}(x^{*},w)$
    \STATE $G \leftarrow \mathrm{UpdateMissGraph}(G,x^{*},y^{*},\Omega,iter)$
    \STATE $S \leftarrow S \cup \{(x^{*},y^{*})\}$, $U \leftarrow U \setminus \{x^{*}\}$, $A \leftarrow A \setminus \{x^{*}\}$
    \STATE $\Omega \leftarrow \mathrm{GetCurrentParetoSet}(S)$
    \STATE $A \leftarrow \mathrm{TCFilter}(A,U,score,iter)$
\ENDFOR
\RETURN $\Omega$
\end{algorithmic}
\end{algorithm}

Figure~\ref{figure2} illustrates the overall framework of PaFeDSE.
Unlike conventional ML-based DSE methods that do not explicitly exploit Pareto-hit and Pareto-miss feedback, PaFeDSE uses HitFormer for Pareto-hit-aware candidate scoring and MissGraph for locality-aware Pareto-miss penalization.
It further reduces redundant candidate processing through TCFilter, which uses inter-iteration consistency to exclude persistently low-priority candidates and retain the remaining candidates as the active candidate pool for acquisition.
These three key components, together with the simulator, form an iterative loop that selects and evaluates promising candidates to refine the observed Pareto set.

Algorithm~\ref{alg:pafedse} provides a more detailed description of the PaFeDSE flow.
Given a design space, PaFeDSE first selects a small initial set using a sensitivity-aware sampling strategy designed to support MissGraph, as described in Section~\ref{sec:missgraph}.
The sampled design points are evaluated with the simulator to obtain their PPA results for the target workload.
Based on these initial evaluations, PaFeDSE constructs the initial observed Pareto set and MissGraph, while initializing the active candidate pool with the remaining unevaluated candidates (Lines~1--5).
After this initialization, PaFeDSE enters an iterative exploration loop over the active candidate pool.

At each iteration, HitFormer is trained using the design points evaluated so far (Line~7).
It then jointly predicts the PPA and Pareto-hit scores of the candidates in the active candidate pool, where the Pareto-hit score indicates how likely a candidate is to become a Pareto hit with respect to the current observed Pareto set.
Using the predicted PPA, the acquisition module calculates the HVI of each active candidate and combines it with the Pareto-hit score from HitFormer and the Pareto-miss penalty from MissGraph (Lines~8--12).
The candidate with the highest acquisition score is selected for the next simulation (Line~13).
After the selected candidate is evaluated by the simulator, PaFeDSE updates the evaluated design point set, the observed Pareto set, and MissGraph, while refining the active candidate pool through TCFilter (Lines~14--18).
This loop continues until the maximum number of iterations is reached, and PaFeDSE returns the final observed Pareto set.

\subsection{Pareto-Hit-Aware Transformer Surrogate}

\begin{figure}[t]
    \centering
    \includegraphics[
        width=\linewidth,
        trim=20 16 20 15,
        clip
    ]{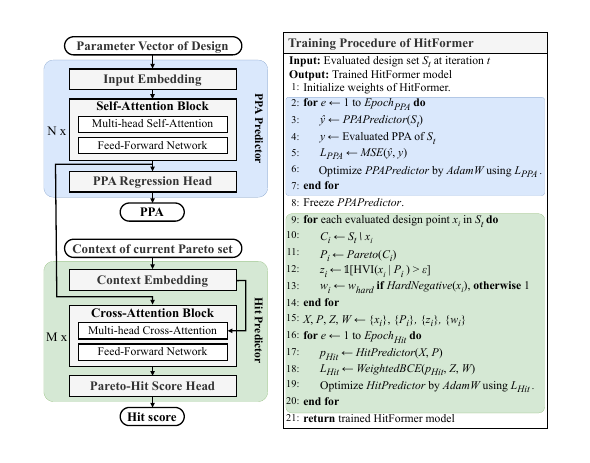}
    \caption{Architecture and training procedure of HitFormer.}
    \Description{Architecture and training procedure of HitFormer, consisting of a self-attention-based PPA predictor and a Pareto-context-aware hit predictor using cross-attention.}
    \label{figure3}
\end{figure}

HitFormer is a Transformer~\cite{attentionisallyouneed}-based surrogate model that augments PPA prediction with Pareto-hit score prediction, enabling candidate selection to consider both the magnitude of predicted HVI and its estimated reliability.
Figure~\ref{figure3} illustrates the architecture and training procedure of HitFormer, which consists of a PPA predictor and a Pareto-hit score predictor.
The PPA predictor embeds the microarchitectural parameters of a candidate design point into tokens and processes them through stacked self-attention blocks, where self-attention captures interactions among architectural parameters.
The PPA predictor then applies a regression head to the resulting candidate representation to predict CPI, power, and area.
In addition to PPA prediction, the Pareto-hit score predictor estimates the likelihood that the candidate will become a Pareto hit.
To make this prediction Pareto-context-aware, the PPA vectors of the current observed Pareto set are embedded into context tokens.
The Pareto-hit score predictor then conditions the candidate representation from the self-attention blocks on these context tokens through multiple cross-attention blocks.
The Pareto-hit score head produces a scalar output and applies a sigmoid function to obtain the final Pareto-hit score in $[0,1]$.
During acquisition, for each candidate design point $x$ with predicted PPA vector $\hat{y}(x)$, the acquisition score is computed as $\mathrm{HVI}(\hat{y}(x)\mid\Omega_t)\cdot p_{\mathrm{hit}}(x)$, where $\Omega_t$ denotes the observed Pareto set at iteration $t$.
The resulting acquisition score favors candidates with large predicted HVI and high Pareto-hit likelihood.

As shown in the training procedure, HitFormer is trained at each DSE iteration $t$ using the evaluated design set $S_{t}$.
The PPA predictor is first trained with the evaluated PPA labels and then frozen.
For the Pareto-hit score predictor, leave-one-out pseudo-DSE samples are constructed by removing each evaluated design point $x_i$ from $S_t$ and forming a context set $C_i=S_t\setminus\{x_i\}$.
Each removed point $x_i$ is labeled as a Pareto hit if adding it to the Pareto set of $C_i$ yields an HVI larger than $\varepsilon$.
We identify \textit{hard negatives} as Pareto-miss samples with high predicted HVI but with actual HVI no greater than $\varepsilon$.
These samples are assigned a larger weight $w_{hard}$ in the binary cross-entropy (BCE) loss to suppress high-HVI false positives.

\subsection{Locality-Aware Pareto-Miss Penalty Graph}
\label{sec:missgraph}

\begin{figure}[t]
    \centering
    \includegraphics[
        width=0.99\linewidth,
        trim=20 9 20 13,
        clip
    ]{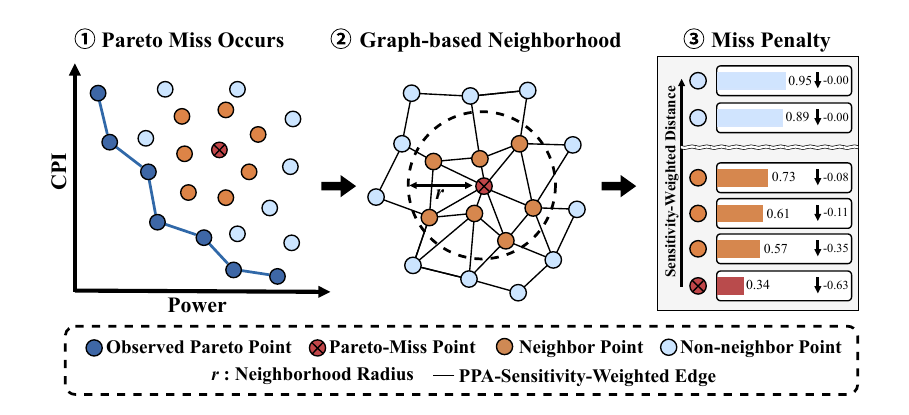}
    \caption{Locality-aware Pareto-miss penalty in MissGraph.}
    \Description{Three-stage illustration showing a Pareto miss, its sensitivity-weighted graph neighborhood, and the assignment of stronger penalties to nearby candidates.}
    \label{figure4}
\end{figure}

MissGraph is a graph-based memory motivated by the locality of microarchitectural design spaces, where nearby configurations often exhibit similar PPA behavior.
Accordingly, MissGraph penalizes candidates close to previously observed Pareto misses, as such candidates are less likely to provide large Pareto improvement.
As shown in Figure~\ref{figure4}, each design point is treated as a graph node, and pairwise distances between nodes serve as edge weights.
MissGraph maintains a non-positive penalty value $q(x)$ for each candidate $x$, initialized to zero.
When a Pareto miss occurs at $x_t$, MissGraph treats candidates whose graph distance from $x_t$ is within the neighborhood radius $r$ as neighbors and subtracts a distance-dependent miss penalty from their penalty values $q(x)$.
Given the base acquisition score from HitFormer, MissGraph applies a graph factor $g(x)=\exp(q(x))$, where $g(x)\in(0,1]$ because $q(x)\leq 0$.
The final acquisition score is computed as $\mathrm{HVI}(\hat{y}(x)\mid\Omega)\cdot p_{\mathrm{hit}}(x,\Omega)\cdot g(x)$.

A naive design-space distance based only on parameter-level differences fails to fully capture locality in the PPA space because different microarchitectural parameters have different impacts on PPA metrics.
Therefore, MissGraph estimates parameter-wise PPA sensitivity from the initial set and incorporates it into the graph distance.
To obtain this sensitivity, \textit{sensitivity-aware initial set sampling} constructs a design pair for each microarchitectural parameter $k$: one design sets $k$ to its lowest level, and the other sets $k$ to its highest level, while the shared values of the remaining parameters are selected by orthogonal sampling~\cite{modse_dac} to ensure broad coverage across the initial set.
After this initial set is evaluated, the normalized differences in CPI, power, and area are aggregated using an $L_1$ norm for each design pair corresponding to parameter $k$, yielding a scalar sensitivity weight $\alpha_k$.
Using these weights, the distance between two design points $x_i$ and $x_j$ is defined as
\begin{equation}
d(x_i,x_j)
=
\frac{
\sum_k \alpha_k \delta_k(x_i,x_j)
}{
\sum_k \alpha_k
},
\end{equation}
where $\delta_k(x_i,x_j)$ denotes the normalized level distance for parameter $k$, computed from the difference between the discrete level indices of $x_i$ and $x_j$.
When a Pareto miss occurs at $x_t$, MissGraph propagates the miss penalty using an RBF kernel~\cite{williams2006gaussian} based on the sensitivity-weighted distance defined above, assigning stronger penalties to candidates closer to $x_t$.
MissGraph also follows a coarse-to-fine schedule that gradually reduces the neighborhood radius while increasing the miss-penalty strength, enabling broad suppression of failed regions in early iterations and fine-grained exploration near the Pareto frontier in later iterations.

\subsection{Temporal-Consistency-Guided Filtering}
\label{sec:tcfilter}

\begin{figure}[t]
    \centering
    \includegraphics[
        width=\linewidth,
        trim=19 12 19 12,
        clip
    ]{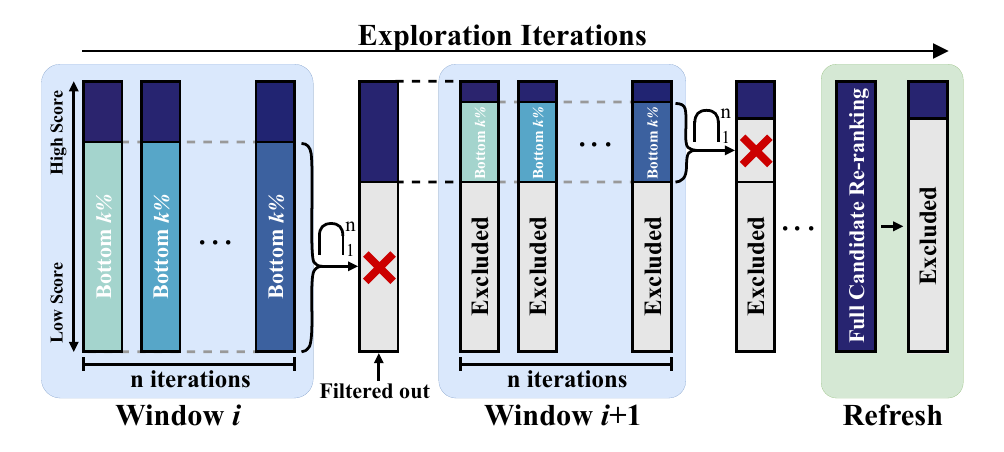}
    \caption{Operating flow of TCFilter across iterations.}
    \Description{Timeline illustrating how TCFilter removes persistently low-scored candidates across iteration windows and periodically restores candidates through full re-ranking.}
    \label{figure5}
\end{figure}

Although Pareto feedback, including the Pareto-hit score from HitFormer and the Pareto-miss penalty from MissGraph, improves the reliability of candidate selection, each iteration still requires scoring a large number of unevaluated candidates to select the next design point.
This exhaustive scoring incurs redundant candidate-processing overhead by repeatedly scoring candidates that are unlikely to be selected, especially when low-scored candidates remain consistently low-priority across adjacent iterations, as observed in Figure~\ref{figure1}(b).
To reduce this overhead, TCFilter maintains an active candidate set $A_t$ and filters out candidates from $A_t$ when they persistently remain low-priority over multiple iterations.

\begin{figure*}[t]
    \centering
    \includegraphics[
        width=0.95\textwidth,
        trim=20 9 20 6,
        clip
    ]{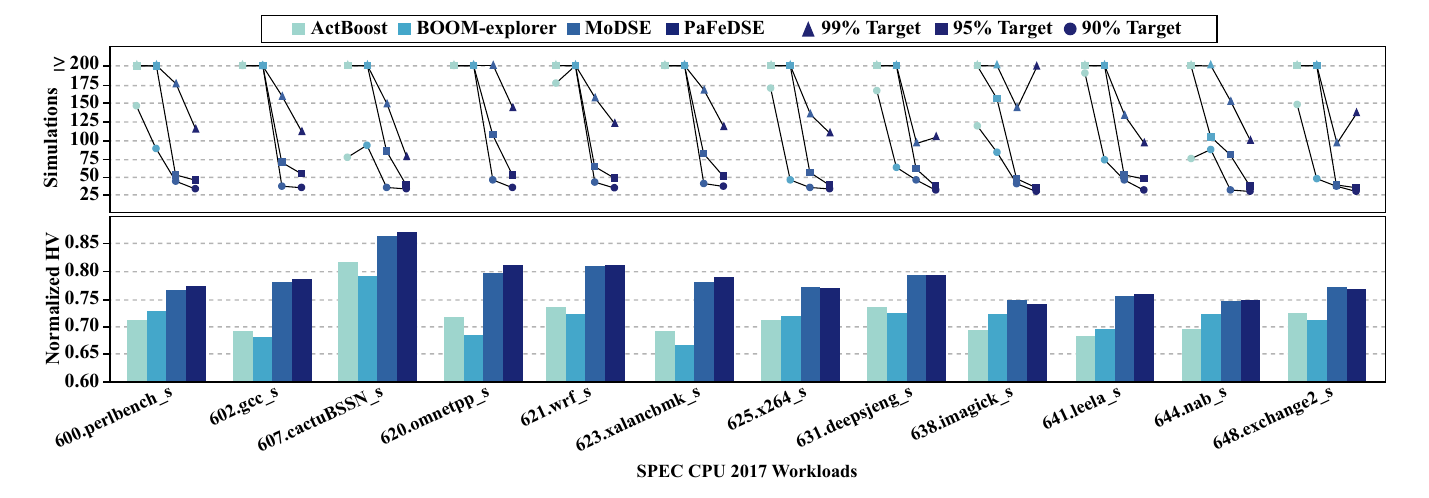}
    \caption{Final normalized HV and simulation counts to reach target HVs defined from the highest final HV among methods.}
    \Description{Comparison of four DSE methods across twelve SPEC CPU 2017 workloads, showing the number of simulations required to reach target HV levels and the final normalized HV.}
    \label{figure6}
\end{figure*}

\begin{table}[t]
  \caption{Microarchitecture design space specification}
  \label{tab:main_design_space}
  \centering
  \footnotesize
  \setlength{\tabcolsep}{2.0pt}
  \renewcommand{\arraystretch}{1.02}
  \begin{tabular}{@{}C{0.05\columnwidth}C{0.45\columnwidth}C{0.46\columnwidth}@{}}
    \toprule
    \textbf{No.} & \textbf{Parameter} & \textbf{Candidate Values} \\
    \midrule

    1 & Pipeline Width & 2, 4, 6, 8 \\
    \midrule
    
    2 & Fetch Buffer Size & 16, 64 \\
    \midrule
    
    3 & Fetch Queue Size & 16, 32, 48 \\
    \midrule

    \multirow{3}{*}{4}
      & BTB/RAS Entries & 1024/16, 2048/24, 4096/32 \\
      & Local Predictor \& History Table Size & 512, 1024, 2048 \\
      & Global \& Choice Predictor Size & 2048, 4096, 8192 \\
    \midrule

    5 & ROB Entries & 64, 96, 128, 160 \\
    \midrule
    
    6 & Instruction Queue Entries & 32, 48, 80 \\
    \midrule

    7 & Physical INT \& FP Registers
      & 160, 192, 256, 304 \\
    \midrule
    
    8 & Load \& Store Queue Entries
      & 16, 32, 40 \\
    \midrule

    9 & Integer ALU Ports & 3, 4, 6 \\
    \midrule
    
    10 & Integer MUL-DIV Ports & 1, 2 \\
    \midrule

    11 & FP ALU Ports/FP MUL-DIV Ports 
       & 1/1, 2/1, 2/2 \\
    \midrule

    12 & L1 ICache Size/Associativity 
       & 16KB/2-way, 32KB/2-way, 64KB/4-way \\
    \midrule

    13 & L1 DCache Size/Associativity 
       & 16KB/2-way, 32KB/2-way, 64KB/4-way \\
    \midrule
    
    14 & L1 DCache MSHRs & 2, 8 \\

    \bottomrule
  \end{tabular}
  \scriptsize
  \scriptsize
  Pipeline width applies to all main pipeline stages. The branch predictor is fixed to TournamentBP.
\end{table}

As illustrated in Figure~\ref{figure5}, TCFilter performs window-based filtering over $n$ consecutive iterations.
At each iteration $t$ in a window, after the final acquisition scores are computed for the active candidate set $A_t$, TCFilter records the bottom $k\%$ lowest-priority candidates as $B_t$.
At the end of the window, TCFilter computes the filtering set as $F_i=\bigcap_{t\in W_i} B_t$, where $W_i$ denotes the iterations in window $i$, and excludes the candidates in $F_i$ from the active candidate set.
Thus, only candidates consistently ranked in the bottom-$k\%$ across the entire window are filtered.
This intersection-based filtering reduces the risk of excluding candidates that are temporarily assigned low priority due to surrogate prediction errors or acquisition-score fluctuations.
To avoid excessive reduction of the active candidate set, TCFilter enforces a minimum active-set size.

Once a candidate is filtered out from the active candidate set, it remains excluded from candidate selection, even though its priority may change as additional simulations improve the surrogate model and update the observed Pareto set.
Therefore, TCFilter periodically refreshes the active candidate set to reconsider previously filtered candidates.
At a refresh iteration, TCFilter recomputes the final acquisition scores for all remaining unevaluated candidates, including previously filtered candidates, and reconstructs the active candidate set from the highest-scored candidates while keeping the active-set size unchanged.
TCFilter then continues window-based filtering on the reconstructed active candidate set.

%%%%%%%%%%%%%%%%%%%%%%%%%%%%%%%%%%%%%%%%%%%%%%%%%%%%%%%%%%%%%%%%%%%%%%%%%%%
%%%%%%%%%%%%%%%%%%%%%%%%%%%%%%%%%%%%%%%%%%%%%%%%%%%%%%%%%%%%%%%%%%%%%%%%%%%

\section{Evaluation}

\subsection{Evaluation Setup}
\label{sec:evaluation_setup}

We use gem5~\cite{gem5} for cycle-accurate simulation and McPAT~\cite{mcpat} for power and area modeling at a 22~nm technology node.
We run SPEC CPU 2017 workloads~\cite{speccpu2017} using SimPoint~\cite{simpoint}, with each SimPoint interval containing 10 million instructions.
Table~\ref{tab:main_design_space} presents the design space used in our evaluation.
The resulting design space contains 3,359,232 design points, making exhaustive evaluation prohibitively expensive in practice.
We fix the L2 cache and DRAM configuration across all design points, using a 2~MiB, 8-way L2 cache with 64~B cache lines and an 8~GB main memory.

We compare PaFeDSE with three ML-based DSE baselines: active-learning-based ActBoost~\cite{actboost}, GP-based BOOM-Explorer~\cite{boomexplorer_iccad}, and MoDSE~\cite{modse_dac}, a state-of-the-art multi-objective DSE method.
We reproduce each baseline as closely as possible following the original papers and available implementations, while fixing the initial set size to 28 for all methods. 
We implement PaFeDSE in Python and HitFormer in PyTorch. 
HitFormer uses an embedding dimension of 96, eight attention heads, and a learning rate of 0.001. 
The PPA predictor consists of four self-attention blocks and is trained for 300 epochs, whereas the Pareto-hit score predictor consists of four cross-attention blocks and is trained for 10 epochs at each DSE iteration.
For TCFilter, we set the window size to $n=8$, the bottom-candidate ratio to $k=80\%$, the refresh period to 16 iterations, and the minimum active-candidate ratio to 1\% of the full design space.
All experiments are conducted on a 36-core Intel Xeon Platinum 8452Y CPU and an NVIDIA RTX 6000 Ada Generation GPU.

\subsection{Overall Pareto Quality and Search Efficiency}

Figure~\ref{figure6} reports the final normalized HV after 200 simulations and the number of simulations required to reach the common target HV levels.
For each workload, all DSE methods use the same HV reference point, obtained by extending the objective-wise worst values by 5\% of their respective ranges.
We normalize PPA to the $[0,1]$ range using the objective-wise best value and the reference point, and compute HV in the normalized objective space.
Since the design space contains 3,359,232 design points, obtaining the true Pareto frontier through exhaustive evaluation is computationally infeasible.
Therefore, we use the best observed HV achieved by the compared methods as a practical reference for evaluating Pareto quality.
This approach provides a common and empirically achievable reference for all methods, enabling a fair comparison of their convergence efficiency under the same simulation budget.
Specifically, the target HVs for each workload are defined as 90\%, 95\%, and 99\% of the highest final HV achieved by any compared method under the 200-simulation budget.
When a method does not reach a target within the budget, we report the budget limit of 200 simulations.

PaFeDSE requires substantially fewer simulations to reach the target HV levels than the prior DSE methods.
At the 99\% target HV level, PaFeDSE requires up to 49.0\% fewer simulations than MoDSE, demonstrating substantially faster convergence toward high-HV regions.
On average, PaFeDSE reaches the 90\%, 95\%, and 99\% target HV levels in 32.9, 43.3, and 118.8 simulations, reducing the simulation count relative to MoDSE by 17.0\%, 34.2\%, and 19.2\%, respectively.
As exploration progresses, HV tends to saturate because the observed Pareto set already dominates a larger portion of the objective space, making differences in final HV less pronounced.
Although further improvement becomes increasingly difficult in this regime, PaFeDSE still achieves the highest average final normalized HV across all workloads under the same 200-simulation budget.
These results show that PaFeDSE combines substantially faster convergence with strong final Pareto quality across the evaluated workloads.

\subsection{Ablation Study}

\begin{figure}[t]
    \centering
    \includegraphics[
        width=\linewidth,
        trim=18 11 20 11,
        clip
    ]{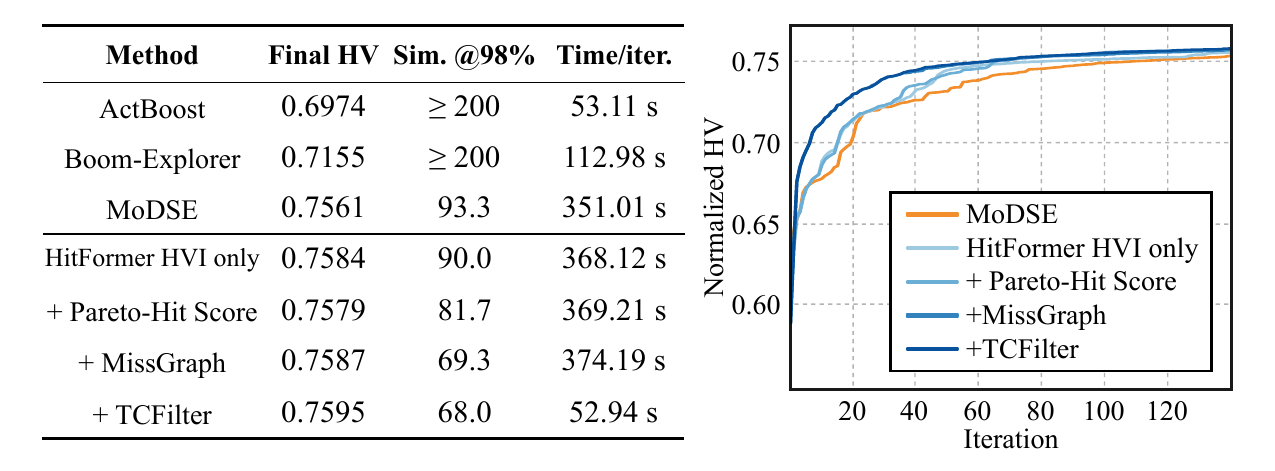}
    \caption{Ablation study of PaFeDSE components.}
    \Description{Table comparing final HV, simulations to target, and exploration time across PaFeDSE variants and baselines, alongside their normalized HV convergence curves.}
    \label{figure7}
\end{figure}

We conduct an ablation study of PaFeDSE on three SPEC CPU 2017 workloads: 600.perlbench\_s, 641.leela\_s, and 644.nab\_s.
Figure~\ref{figure7} reports component-level results averaged across these workloads under a 200-simulation budget, including final HV, simulations required to reach the 98\% target HV, average exploration time per iteration, and the normalized HV curves.
We use the 98\% target HV to compare component contributions in a high-HV region without excessive sensitivity to the last few iterations.

We define exploration time as non-simulation overhead, including model training, model inference, and acquisition scoring.
All methods are measured on the same hardware platform using execution backends consistent with their original papers and available implementations.
ActBoost and MoDSE use CPU-based learning and acquisition components, while BOOM-Explorer runs on the GPU following its BoTorch-based acquisition implementation.
PaFeDSE uses the GPU for HitFormer training and inference and the CPU for acquisition scoring, matching the CPU-based HVI scoring backend of MoDSE for a fair comparison.

Starting from the HVI-only version of HitFormer, we incrementally add the Pareto-hit score, MissGraph, and TCFilter to evaluate the effect of each component.
Adding the Pareto-hit score reduces the average number of simulations required to reach the 98\% target HV from 90.0 to 81.7, and MissGraph further reduces it to 69.3.
This shows that Pareto-hit and Pareto-miss feedback improve the simulation efficiency of PaFeDSE.
TCFilter reduces the average exploration time per iteration from 374.19~s to 52.94~s, achieving a 7.1$\times$ reduction without degrading final HV.
The HV curves show that progressively adding PaFeDSE components substantially accelerates HV improvement and yields higher HV than MoDSE.

\subsection{Detailed Analysis of PaFeDSE}

\begin{figure}[t]
    \centering
    \includegraphics[
        width=\linewidth,
        trim=19 12 19 14,
        clip
    ]{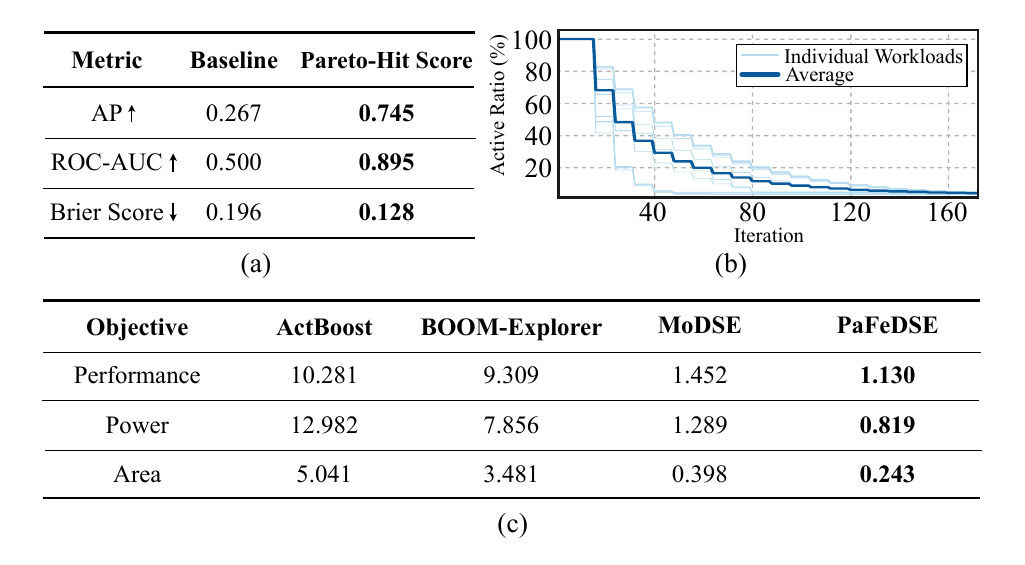}
    \caption{Detailed analysis of PaFeDSE.
    (a) Pareto-hit prediction performance.
    (b) Active candidate ratio across iterations.
    (c) Average MAPE of PPA predictions across methods.}
    \Description{Three analyses showing Pareto-hit prediction metrics, the decreasing active candidate ratio over iterations, and PPA prediction errors across DSE methods.}
    \label{figure8}
\end{figure}

Figure~\ref{figure8} presents detailed analyses of Pareto-hit prediction quality, active candidate ratio, and PPA prediction accuracy.
We evaluate the Pareto-hit score using Average Precision (AP)~\cite{davis2006prroc}, the area under the receiver operating characteristic curve (ROC-AUC)~\cite{hanley1982roc}, and the Brier score~\cite{brier1950}.
AP measures whether true Pareto-hit candidates are ranked near the top, while ROC-AUC measures how well Pareto-hit and Pareto-miss candidates are separated across all possible classification thresholds.
A ROC-AUC of 0.5 corresponds to random scoring, and a value closer to 1.0 indicates better separation.
The Brier score measures the mean squared error between the predicted Pareto-hit score and the binary Pareto-hit label.
We evaluate PPA prediction accuracy using Mean Absolute Percentage Error (MAPE)~\cite{hyndman2006accuracy}, which measures the average percentage error between predicted and true PPA values.
For each method, we compute MAPE over the design points selected during its DSE process by comparing the PPA predicted before simulation with the simulated PPA of each selected point.

Figure~\ref{figure8}(a) evaluates the quality of the Pareto-hit score on candidates selected during the actual DSE runs.
The baseline uses random ranking for AP and ROC-AUC and assigns the observed Pareto-hit rate as a constant probability for the Brier score.
The Pareto-hit score substantially outperforms the baseline across all metrics, showing that HitFormer can reliably distinguish Pareto-hit candidates from Pareto-miss candidates. 
Figure~\ref{figure8}(b) shows the active candidate ratio, defined as $|A_t|/|U_t|$.
The average active ratio steadily decreases and reaches as low as 1.82\%, indicating that TCFilter filters out the vast majority of candidates. 
Figure~\ref{figure8}(c) compares PPA prediction MAPE averaged across the workloads.
PaFeDSE achieves an average PPA MAPE of 0.731\%, reducing the average prediction error by 30.2\% compared with MoDSE.

\subsection{Time Analysis of PaFeDSE}

\begin{figure}[t]
    \centering
    \includegraphics[
        width=\linewidth,
        trim=20 13 20 12,
        clip
    ]{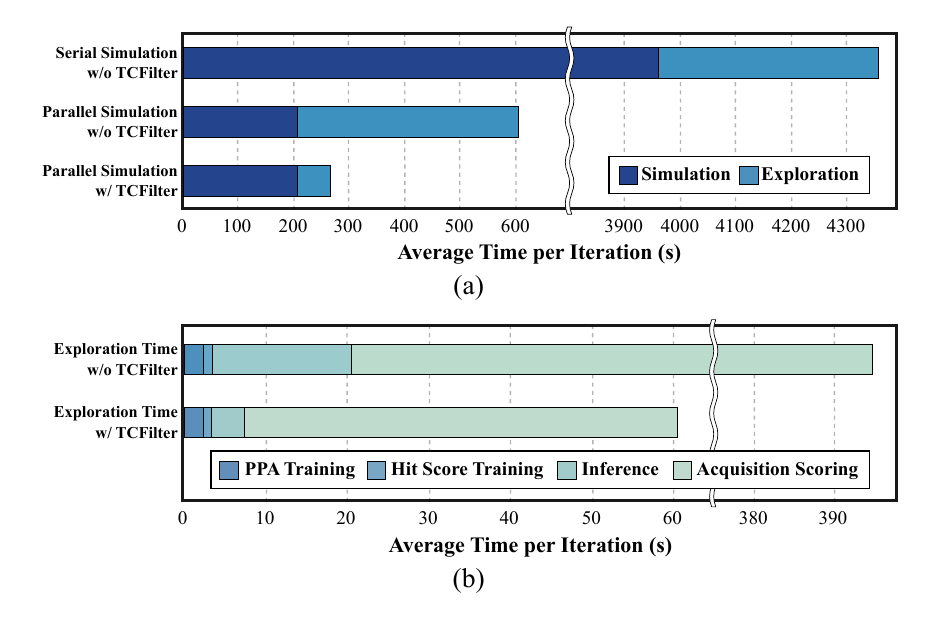}
    \caption{Per-iteration end-to-end time analysis of PaFeDSE on 600.perlbench\_s. (a) Total execution time per iteration. (b) Exploration time breakdown.}
    \Description{Stacked bar charts comparing evaluation and exploration times under serial and parallel simulation, with and without TCFilter, and breaking down exploration time into training, inference, and acquisition scoring.}
    \label{figure9}
\end{figure}

To analyze the execution time of PaFeDSE, we measure the average per-iteration time on 600.perlbench\_s using the CPU and GPU platform described in Section~\ref{sec:evaluation_setup}.
Each DSE iteration consists of two main stages: design point evaluation and exploration.
In practical hardware design flows, design evaluation may involve cycle-accurate architectural simulation, RTL simulation, post-layout simulation, or hardware emulation, depending on the design stage.
Although evaluation time varies substantially across these approaches, design evaluation often incurs significant computational cost and can dominate the overall DSE execution time.
In this study, we use gem5~\cite{gem5} and McPAT~\cite{mcpat}, both of which are included in the measured simulation time.
To accelerate the simulation, we employ SimPoint~\cite{simpoint} to select representative execution intervals instead of simulating entire workloads.
While gem5 simulation typically runs serially on a single host CPU core, checkpoints for SimPoint intervals allow these intervals to be simulated independently and in parallel across multiple host CPU cores.
We therefore consider both serial and parallel SimPoint execution to examine the impact of simulation parallelism on the DSE execution time.

Figure~\ref{figure9}(a) presents the average per-iteration execution time, including evaluation and exploration.
Under serial SimPoint execution, gem5 simulation and McPAT estimation require 3919.9~s and 42.8~s, respectively, resulting in a total evaluation time of 3962.7~s.
Parallel SimPoint execution reduces the gem5 simulation time to 195.1~s and the McPAT estimation time to 12.8~s, reducing the total evaluation time to 207.9~s.
Including exploration time, the overall execution time decreases from 4357.5~s to 602.7~s per iteration without TCFilter.
By reducing the active candidate pool, TCFilter decreases the average exploration time per iteration from 394.8~s to 60.4~s.
Consequently, the total execution time per iteration under parallel evaluation further decreases from 602.7~s to 268.3~s.
These results indicate that although design evaluation typically dominates DSE execution time, its cost varies depending on the evaluation method and degree of parallelism, making exploration overhead significant as evaluation time decreases.

Figure~\ref{figure9}(b) further breaks down exploration time into PPA training, Pareto-hit score training, inference, and acquisition scoring.
Acquisition scoring includes all remaining measured exploration operations beyond model training and inference, such as HVI computation, integration of Pareto-hit scores and MissGraph penalties, and candidate selection.
In particular, computing three-dimensional HVI for each active candidate against the current observed Pareto set incurs substantial computational cost, making acquisition scoring the dominant component of exploration time.
As shown in Figure~\ref{figure9}(b), TCFilter substantially reduces both inference and acquisition scoring time by reducing the number of candidates requiring these operations.
Meanwhile, the combined training time remains nearly unchanged, indicating that TCFilter primarily reduces redundant candidate processing rather than model training costs.
Consequently, TCFilter reduces the average exploration time per iteration from 394.8~s to 60.4~s on 600.perlbench\_s.

\section{Related Work}

Machine-learning-based microarchitecture DSE has been widely studied to reduce expensive design evaluations.
ActBoost~\cite{actboost} uses active learning, BOOM-Explorer~\cite{boomexplorer_iccad} combines GP-based surrogate modeling with EHVI-based acquisition, and MoDSE~\cite{modse_dac} improves multi-objective CPU DSE with Pareto-aware surrogate training, HVI-based selection, and uniformity-aware candidate selection.
Recent learning-based approaches also exploit graph or ranking-based representations~\cite{rankdse} and attention mechanisms~\cite{attentiondse} to improve exploration efficiency.
Bottleneck-analysis-based methods such as ArchExplorer~\cite{archexplorer} improve interpretability by identifying architectural bottlenecks during exploration.
In contrast, PaFeDSE explicitly leverages Pareto-hit and Pareto-miss outcomes to guide candidate selection and exploits temporal consistency to reduce redundant surrogate inference and acquisition scoring.

\section{Conclusion}

This paper presented PaFeDSE, a Pareto-feedback-guided framework for efficient CPU microarchitecture DSE.
PaFeDSE combines HitFormer, MissGraph, and TCFilter to improve Pareto-oriented candidate selection and reduce redundant acquisition scoring.
Across SPEC CPU 2017 workloads, PaFeDSE requires up to 49.0\% fewer simulations to reach high-HV regions and reduces PPA prediction error by 30.2\% over the strongest prior baseline, while adaptive filtering achieves a 7.1$\times$ reduction in exploration overhead.
PaFeDSE also achieves the highest average final normalized HV under the same simulation budget, demonstrating its effectiveness in identifying high-quality Pareto solutions.
These results show that exploiting Pareto feedback and temporal candidate consistency is an effective direction for improving both simulation and candidate-processing efficiency in microarchitecture DSE.

\begin{acks}
This work was supported by Samsung Electronics Co., Ltd(IO251211-14291-01).
\end{acks}

\bibliographystyle{ACM-Reference-Format}
\bibliography{references}

\end{document}